\documentclass[aps,prl,reprint,superscriptaddress]{revtex4-2}

\usepackage{graphicx}% Include figure files
\usepackage{dcolumn}% Align table columns on decimal point
\usepackage{bm}% bold math
\usepackage[normalem]{ulem}
\usepackage{amsmath,amssymb,amsfonts}%
\usepackage{mathrsfs}%
\usepackage{textcomp}%
\usepackage{babel} %%% \glqq \grqq
\usepackage{svg} %%% \includesvg
\usepackage{tikz}
\usetikzlibrary{math,calc,arrows.meta}
\usetikzlibrary{decorations.pathreplacing}  % Required for braces
\usepackage{adjustbox}

\newcommand{\uproman}[1]{\uppercase\expandafter{\romannumeral#1}}

\newcommand{\dt}{{\mathrm{d}}}

\newcommand{\imagu}{\mathrm{i}}
\newcommand{\eulere}{\mathrm{e}}

\begin{document}

\title{Evidence for dynamical chiral condensate in high-energy heavy ion collisions}

\author{Tobias Bruschke}
\email{tobias.bruschke@desy.de}
\affiliation{\uproman{2}. Institut für Theoretische Physik, Universität Hamburg, 22761 Hamburg, Germany}
\affiliation{Theoretisch-Physikalisches Institut, Friedrich-Schiller Universität, 07743 Jena, Germany}

\author{Andreas Kirchner}
\email{andreas.kirchner@duke.edu}
\affiliation{Department of Physics, Duke University, Durham, NC 27708, USA }

\author{Stefan Floerchinger}
\email{stefan.floerchinger@uni-jena.de}
\affiliation{Theoretisch-Physikalisches Institut, Friedrich-Schiller Universität, 07743 Jena, Germany}


\begin{abstract}
Quantum chromodynamics with light quarks features an approximate global symmetry, known as chiral symmetry, that is believed to be spontaneously broken by the vacuum expectation value of a scalar and isoscalar composite field, in addition to a small explicit breaking due to finite quark masses. For a high enough temperature, as achieved in the early universe or the fireball created by a high-energy heavy ion collision, this symmetry is expected to be restored. We show theoretically that a coherent deviation of the corresponding quantum field from its usual vacuum expectation value on the freeze-out hypersurface of a heavy-ion collision leads, after resonance decays, to a characteristic contribution to the transverse momentum spectrum of charged pions, in the very soft regime, consistent with experimental data from the Relativistic Heavy Ion Collider and the Large Hadron Collider. Taken together, the experimental data with the new theoretical results provide compelling support for the existence of a chiral condensation mechanism with partial restoration of chiral symmetry at high temperature. 
\end{abstract}

\maketitle

%\tableofcontents

\paragraph*{Introduction}\label{sec:introduction}

Quantum chromodynamics (QCD) is the accepted theory of the strong nuclear force, and an integral part of the standard model of elementary particle physics. In the low-energy regime, the gauge coupling grows large, which leads to a number of interesting phenomena, like color confinement, but also makes theoretical investigations difficult. In this situation, like often in physics, symmetries are an important guiding principle. In the limit where the up and down quarks are (almost) massless, the Lagrangian of QCD has an (approximate) global chiral symmetry $\text{SU}(2)_L \times \text{SU}(2)_R \times \text{U}(1)$. If this were an unbroken symmetry, it would imply, for example, that mesons and baryons had partners with opposite parity, but otherwise equal properties, like mass. It is widely believed that this symmetry is broken spontaneously down to the subgroup $\text{SU}(2)_V \times \text{U}(1)$ by the vacuum expectation value of a composite field that transforms as scalar with respect to Lorentz transformations, and as a singlet (``isoscalar'') with respect to the unbroken isospin group $\text{SU}(2)_V$ \citep{Donoghue:1992dd}. This spontaneous breaking of the chiral symmetry leads to a lift of the mass degeneracy of parity partners, and to the appearance of three almost massless pseudo-Goldstone bosons \citep{NambuJona-Lasinio_1961, Goldstone_1961, GoldstoneEtAl_1962}, the charged and neutral pions, $\pi^\pm$ and $\pi^0$. The symmetry-breaking scalar field is denoted as $\sigma$ with vacuum expectation value equal to the pion decay constant, $\langle \sigma \rangle = f_\pi \approx 92.3 \text{ MeV}$ \cite{Navasothers_2024}. Excitations in this field around the vacuum expectation value are identified with the $\sigma / f_0(500)$ resonance \cite{Pelaez_2016}.

While this picture is certainly very compelling, direct experimental evidence for the chiral condensate and the corresponding spontaneous symmetry breaking mechanism is missing so far. Of course, a vacuum expectation value as such cannot be detected. However, a quantum field that is displaced away from its vacuum expectation value in some region of spacetime defines a coherent state \cite{Glauber_1963}, and could be detected experimentally. Examples for coherent quantum fields arise, besides quantum optics, frequently in condensed matter \citep{Schleich2001,Altland_Simons_2010}. A particularly illustrative example is provided by Bose-Einstein condensates in ultra-cold atomic quantum gases. Here, it is a non-relativistic matter field that can be seen as being displaced from its (vanishing) vacuum value, and the associated spontaneous symmetry breaking leads, for example, to superfluidity. Experimentally, the coherent field or condensate was first detected via the distribution of its constituent particles in momentum space, measured by time-of-flight after release from a trap \cite{CornellWieman_2002, Ketterle_2002}. A peak at low momenta, corresponding to the squared Fourier transform of the condensate field, has been seen there at low enough temperatures, in addition to a thermal distribution of non-condensed atoms \citep{AndersonEtAl_1995, DavisEtAl_1995, BradleyEtAl_1995}. We will discuss below that a coherent field deviation from the chiral condensate vacuum expectation value has a very similar experimental signature. 

For QCD matter, there is a transition temperature above which the chiral condensate vanishes or becomes very small, such that chiral symmetry is at least partially restored \cite{PisarskiWilczek_1984}. For realistic quark masses, this transition is a continuous crossover \cite{AokiEtAl_2006}, at a temperature of about $155$ MeV \cite{AokiEtAl_2006a, AokiEtAl_2009, HotQCDCollaborationEtAl_2014}. This also implies that when QCD matter undergoes a transition from the high-temperature to a low-temperature phase, this is accompanied with a change in the chiral condensate. When the temperature changes very slowly on the relevant timescales of QCD, the chiral condensate has enough time to adapt and should follow the same function of temperature that is calculated for a global equilibrium state. On the other side, when the temperature changes quickly, it is possible that the chiral condensate lags behind, and has essentially still a high-temperature value while the temperature has already dropped below the crossover temperature. The displaced chiral order parameter field value is then showing up as a coherent $\sigma/f_0(500)$ field. It is this scenario of a partially restored chiral condensate (PRCC) at freeze-out that we investigate below in further detail. 

Let us note here that a different, albeit related scenario has been discussed previously -- a local misalignment of the chiral condensate in isospin space from the $\sigma$ to the $\vec\pi$ directions, also known as disoriented chiral condensate \cite{Anselm_1989, RajagopalWilczek_1993, Bjorken_1997, Amelino-CameliaEtAl_1997, MohantySerreau_2005,Blaizot:1992at,Anselm:1991pi}. More recently, a possible formation of a Bose-Einstein condensate of pions was theoretically explored \citep{BegunFlorkowski_2015}, as well as the phenomenological influence of critical fluctuations in the chiral order parameter field \citep{FlorioEtAl_2025,Florio:2025lvu}. Indirect experimental signals for chiral symmetry restoration are being explored in ratios of particle yields of chiral partners, strange to non-strange particles \citep{CassingEtAl_2016, PalmeseEtAl_2016, MoreauEtAl_2017}, as well as dilepton spectra \citep{RappEtAl_2010, RappvanHees_2016, SungEtAl_2021, ZhouEtAl_2024,Seck:2020qbx,Savchuk:2022aev}. 

Additional studies on the impact of the $\sigma$ and other resonances in the statistical hadronization model and the resulting change in the pion yield have been conducted in \cite{Broniowski:2015oha,Andronic:2008gu}.

\paragraph*{High-energy heavy ion collisions and fluid-dynamic description}

A fireball of QCD matter at high temperature can be created with high-energy nuclear collisions, as conducted at the Relativistic Heavy Ion Collider (RHIC) \citep{Arsene_2005, BackPHOBOSCollaboration_2005, PHENIXCollaborationAdcox_2005, STARCollaborationAdams_2005} or at the Large Hadron Collider (LHC) \citep{ALICE:2022wpn,Schukraft:2013wba,Andronic:2014zha,Foka:2016vta}. The standard theoretical description of the soft dynamics of a heavy ion collision at high collision energy is in terms of relativistic fluid dynamics \citep{Muller:2013dea,Busza:2018rrf,Romatschke:2017ejr,Soloviev:2021lhs,Jaiswal:2016hex,Shen:2020mgh}. 
This is based on the observation that the dynamics of strongly interacting quantum fields on relatively large length and time scales, i.\ e.\ for slow enough evolution, can be described effectively with thermodynamic concepts employed locally. Fluid-dynamic approximations that are based on this rationale describe many strongly interacting quantum systems and find applications from condensed matter to astrophysics \cite{Landau1987Fluid,Rezzolla:2013dea}. The equations of motion are obtained from the conservation laws for energy and momentum, as well as baryon number and other conserved charges (like electric charge or heavy quark numbers). These get supplemented by additional equations of motion for the dissipative corrections, which are postulated in a more phenomenological way (see e.g. \ \citep{ISRAEL1979341}). The final part of the description, characterizing the fluid and its microscopic dynamics, are the equation of state (taken from lattice QCD calculations \citep{HotQCDCollaborationEtAl_2014}) and its transport properties, such as the viscosities.

For the fluid-dynamic description of the collision, we employ the same setup as \cite{LuEtAl_2025}, where the fluid-dynamic equations of motion are being solved using the Fluid\textit{u}M framework \citep{Floerchinger:2013hza,Floerchinger:2014fta,FloerchingerEtAl_2019} in conjunction with initial conditions obtained from the T\raisebox{-.5ex}{R}ENTo model \citep{Moreland:2014oya,Moreland:2019szz}. During the evolution, the QGP fireball expands, dilutes, and cools down until hadrons and their resonances start to propagate independently; their momentum distributions freeze out.

The momentum space distribution of these hadrons can be calculated from the fluid fields on the freeze-out hypersurface, corresponding to the spacetime manifold with constant temperature $T_\mathrm{fo}$. We use the code \textsc{FastReso} \cite{Mazeliauskas:2018irt} which also allows for the inclusion of resonance decays and hadronic rescatterings (also known as partial chemical equilibrium \citep{Mazeliauskas:2018irt, Kirchner:2023fsj,PhysRevC.101.014910,Bebie:1991ij}). The final result are transverse momentum space distributions of identified charged particles, that can be compared to experimental data \citep{Bernhard:2018hnz,JETSCAPE:2023nuf,Nijs:2020roc,Vermunt:2023zsk,Paquet:2023rfd}. The values for all appearing parameters in this model are fixed using the calibration conducted in \citep{LuEtAl_2025}.

%On the freeze-out hypersurface, corresponding to the spacetime manifold where the temperature has the above value, hadronic particles and resonances are assumed to start freely propagating (apart from strong resonance decays that still take place)s. Their momentum space distribution can be calculated from the fluid dynamic description. We use the code \textsc{FastReso} to determine the feed-down from resonance decays to the observed hadrons. The final result are transverse momentum space distributions of identified charged particles, that can be compared to experimental data \todo{[refs to papers with theory-experiment comparison]}.

\paragraph*{Coherent Particle Production from Classical Condensate Dynamics}

\begin{figure}
   \centering
        \begin{tikzpicture}[scale=1.4]
            \tikzmath{
                \fpisq1=1;
                \fpisq2=-0.3;
                \b=0.7;
                \a1={-2*\b*\fpisq1};
                \a2={-2*\b*\fpisq2};
                \scalex=2;
                \radius = 0.15;
                \labeldcc=-1;
                function Vvac(\x) {
                    return \a1*\x*\x+\b*\x*\x*\x*\x;
                };
                function Vtemp(\x) {
                    return \a2*\x*\x+\b*\x*\x*\x*\x;
                };            
            }
    
            \draw[->] (-0.6*\scalex,0) -- (1.7*\scalex,0) node[label=right:$\sigma$] {};
            \draw[->] (0,-1) -- (0,2) node[label=above:$V(\sigma)$] {};
            
            \draw[thick, domain=-0.5:1.5, samples=30, variable=\x] plot ({\scalex*\x,Vvac(\x)}) node[label=above:$T<T_{\text{chiral}}$,xshift=10pt] {};
            \draw[thick, domain=-0.5:1.2, samples=30, variable=\x] plot ({\scalex*\x,Vtemp(\x)}) node[label=above:$T>T_{\text{chiral}}$,xshift=10pt] {};
    
            \draw[thick, red, ->,>=latex, domain=0.05:{sqrt(\fpisq1)-0.05}, samples=15, variable=\x] plot ({\scalex*\x},{Vvac(\x)+0.7*\radius});
            \node[red, below,yshift=-5pt] at ({\scalex*0.5*sqrt(\fpisq1)}, {Vvac(0.5*sqrt(\fpisq1))+0.7*\radius}) {3};
            
            \draw[thick, red, ->,>=latex, domain={sqrt(\fpisq1)-0.05}:0.05, samples=15, variable=\x] plot ({\scalex*\x},{Vtemp(\x)+1.3*\radius});
            \node[red, above] at ({\scalex*0.5*sqrt(\fpisq1)}, {Vtemp(0.5*sqrt(\fpisq1))+2.5*\radius}) {2};

            \draw[thick, red, ->, >=latex] ({\scalex*sqrt(\fpisq1)}, {Vvac(sqrt(\fpisq1))}) .. controls +(0.4,0.5) and +(0.4,-0.5) .. ({\scalex*sqrt(\fpisq1)}, {Vtemp(sqrt(\fpisq1))});
            \node[red, right] at ({\scalex*sqrt(\fpisq1)+0.3}, {0.3*Vtemp(sqrt(\fpisq1))}) {1};
            \draw ({\scalex*sqrt(\fpisq1)}, -0.1) -- ({\scalex*sqrt(\fpisq1)}, 0.1) node[label=above:$f_\pi$,yshift=-6pt] {};
    
            \foreach \n/\op in {0/0.05, 1/0.10, 2/0.15, 3/0.20, 4/0.25, 5/0.50} {
                \pgfmathsetmacro{\x}{\n*0.8*\radius/\scalex};
                \draw[fill, line width=0, opacity=\op] ({\scalex*\x},{Vvac(\x)+\radius}) circle ({\radius});
            }
            \foreach \n/\op in {0/0.05, 1/0.10, 2/0.15, 3/0.20, 4/0.50} {
                \pgfmathsetmacro{\x}{sqrt(\fpisq1)-\n*0.3*\radius/\scalex};
                \draw[fill, line width=0, opacity=\op] ({\scalex*\x},{Vtemp(\x)+\radius}) circle ({\radius});
            }
            \foreach \n/\op in {0/0.50, 1/0.15, -1/0.15, 2/0.10, -2/0.10, 3/0.05, -3/0.05} {
                \pgfmathsetmacro{\x}{sqrt(\fpisq1)+\n*0.7*\radius/\scalex};
                \draw[fill, line width=0, opacity=\op] ({\scalex*\x},{Vvac(\x)+\radius}) circle ({\radius});
            }
    
            \pgfmathsetmacro{\xsigmamax}{5*0.8*\radius/\scalex};
            \draw[decorate, decoration={brace, mirror, amplitude=5pt}] ({\scalex*\xsigmamax},\labeldcc-0.05) -- ({\scalex*sqrt(\fpisq1)},\labeldcc-0.05) node[midway,yshift=-12pt] {$\Delta\sigma_{\text{coherent}}$};
            \draw[<->,>=latex] ({\scalex*\xsigmamax},\labeldcc) -- ({\scalex*sqrt(\fpisq1)},\labeldcc);
            \draw[dashed] ({\scalex*\xsigmamax},0) -- ({\scalex*\xsigmamax},\labeldcc);
            \draw[dashed] ({\scalex*sqrt(\fpisq1)},{Vvac(sqrt(\fpisq1))}) -- ({\scalex*sqrt(\fpisq1)},\labeldcc);
        \end{tikzpicture}
    \caption{Potential of the chiral condensate for different temperature regimes. Before the collision, the chiral condensate is at its vacuum value $\sigma = f_\pi$. At temperatures above the chiral crossover, the effective potential is modified such that it has its minimum at vanishing or very small $\sigma$ (1), leading to a relaxation of the condensate toward the new minimum (2). When the fireball expands and cools down, the vacuum form of the potential is restored for low enough temperatures. If the cooling takes place quickly, the chiral condensate can lag behind and still deviate from the vacuum expectation value at the moment of freeze-out (3). This difference has the physical significance of a coherent field, finally resulting in an extra production mechanism for low momentum pions.}
    \label{fig:ccquench}
\end{figure}
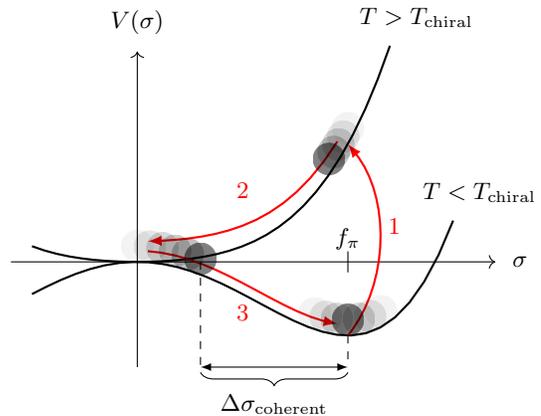

Let us now discuss how a displaced chiral order parameter field arises during a heavy ion collision and its contributions to the final state particle spectra. If chiral symmetry is (partially) restored in the high-temperature regime, this corresponds to a substantial displacement of the chiral order parameter field inside the fireball from the vacuum expectation value on the outside, as illustrated in Fig.~\ref{fig:ccquench}: Before the collision, the condensate is at its vacuum value $\sigma = f_\pi$. At temperatures above the chiral crossover, the effective potential is modified such that it has its minimum at vanishing or very small $\sigma$ (1 in Fig.~\ref{fig:ccquench}), leading to a relaxation of the condensate toward the new minimum (2 in Fig.~\ref{fig:ccquench}). When the fireball expands and cools down, the vacuum form of the potential is restored for low enough temperatures. If the cooling takes place quickly, the chiral condensate can lag behind and still deviate from the vacuum expectation value at the moment of freeze-out (3 in Fig.~\ref{fig:ccquench}). Neglecting for a moment the decay width of the $\sigma/f_0(500)$, this deviation can be modeled as a coherent quantum state with quantum field expectation value $\sigma(t, \mathbf{x})$ that fulfills the equation of motion $[(\partial/\partial t)^2- \boldsymbol{\nabla}^2+m^2][\sigma(t, \mathbf{x})-f_\pi]=J(t, \mathbf{x})$. Here, the real source function $J(t, \mathbf{x})$ is due to interactions with other fields and also encodes deviations in the effective potential from the small amplitude vacuum form $(1/2) m^2 (\sigma-f_\pi)^2$. In this formalism, one can determine the momentum distribution of produced $\sigma/f_0(500)$ particles, once the source $J(t, \mathbf{x})$ has ceased, to be \cite{fsu_mods_00027527, Amelino-CameliaEtAl_1997}
\begin{equation}
    E_\mathbf{p}\frac{\dt N}{\dt^3p}=\frac{1}{2}\frac{J(\mathbf{p})^* J(\mathbf{p})}{(2\pi)^3}\,,
    \label{eq:SpectrumFromSource}
\end{equation}
where $J(\mathbf{p})=\int\dt t\,\dt^3x\,J(t, \mathbf{x})\eulere^{\imagu E_\mathbf{p} t-\imagu \mathbf{p} \mathbf{x}}$ is the Fourier-transformed source function evaluated on-shell, with $E_\mathbf{p} = \sqrt{\mathbf{p}^2+m^2}$. 

The construction of $J(t,\mathbf{x})$ from first principles is challenging, but not strictly necessary, since the relevant information is only required on the freeze-out hypersurface. The source $J(\mathbf{p})$ can then be expressed as integral of a current $j^\mu(t, \mathbf{x}, \mathbf{p})$ over the freeze-out hypersurface,
\begin{equation}
    J(\mathbf{p})= \int_{\Sigma}\dt\Sigma^\mu j_\mu(t, \mathbf{x}, \mathbf{p}),    
    \label{eq:SourceFT_Freezeout}
\end{equation}
with the current depending on the field expectation value,
\begin{equation}
j_\mu(t, \mathbf{x}, \mathbf{p}) = \left(\partial_\mu \sigma(t, \mathbf{x}) + i p_\mu [\sigma(t, \mathbf{x}) - f_\pi] \right) e^{i E_\mathbf{p} t - i \mathbf{p} \mathbf{x}}.
\label{eq:defCurrent}
\end{equation}
This current is conserved, $\nabla_\mu j^\mu(t, \mathbf{x}, \mathbf{p})= 0$, allowing for the deformation of the hypersurface $\Sigma$, in spacetime regions with $J(t, \mathbf{x}) = 0$. In particular, this allows us to calculate the particle production from a coherent deviation of the chiral condensate from its vacuum value at the moment of freeze-out. An identical formalism was already developed in the context of the color class condensate \citep{GelisVenugopalan_2006, Gelis_2013, Gelis_2019}, a hypothesized state of macroscopically occupied gluon modes.

Equation \eqref{eq:SpectrumFromSource}, together with \eqref{eq:SourceFT_Freezeout} and \eqref{eq:defCurrent} should be contrasted with the Cooper-Frye freeze-out formula \citep{CooperFrye_1974}
\begin{equation}
    E_\mathbf{p}\frac{\dt N}{\dt^3p}= \int_\Sigma \mathrm{d}\Sigma^\mu p_\mu f(x,p),
\end{equation}
commonly used to translate fluid fields to particle momentum distributions. The Cooper-Frye integral expresses the final particle spectrum as a sum of Fermi-Dirac- or Bose-Einstein-distributed spectra from each fluid cell on the freeze-out surface. In contrast to that, the particle spectrum from a coherent state computed here in terms of a Fourier transform on a general spacetime hypersurface is non-local, and contributions to the particle distribution can not be uniquely associated with individual points on the freeze-out surface.

\begin{figure}
    \centering
    \includegraphics[width=0.9\linewidth]{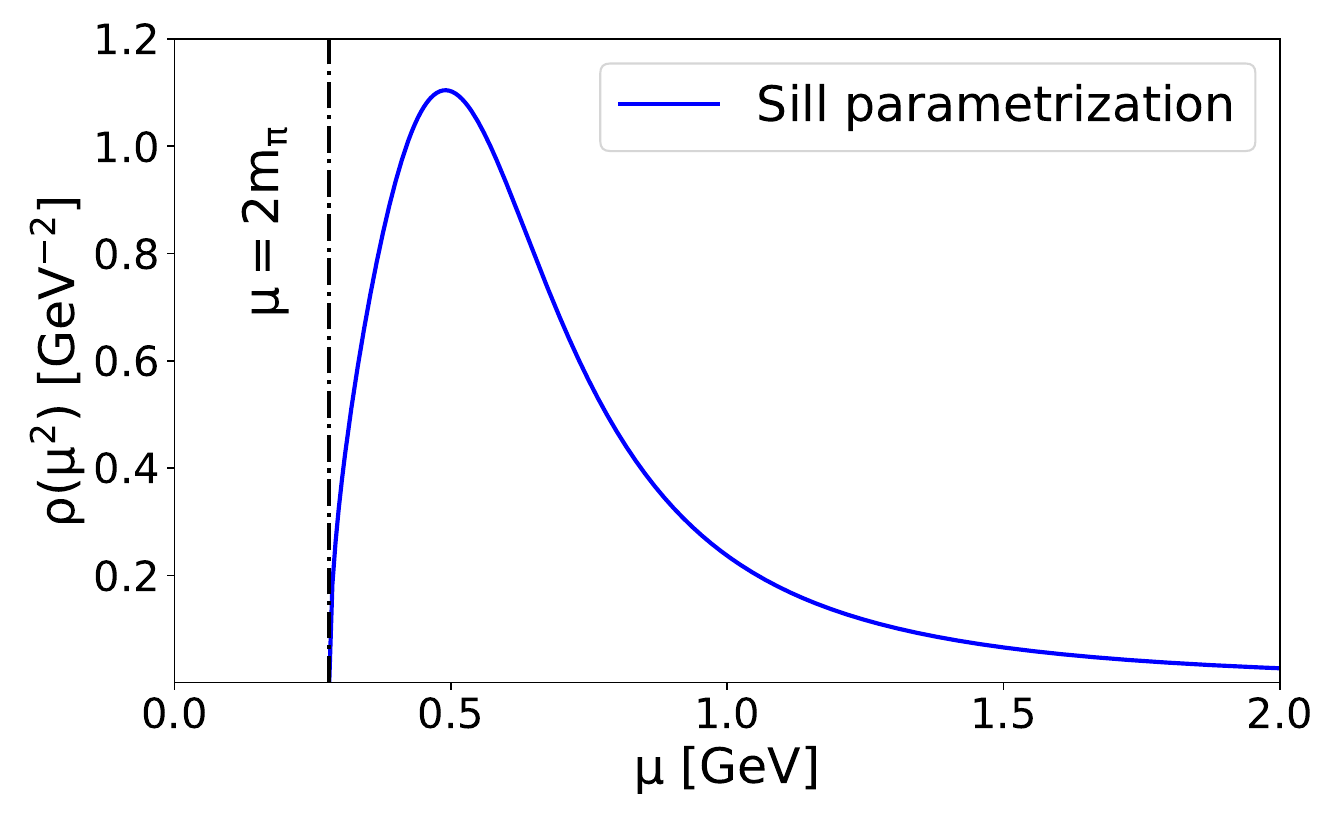}
    \caption{Spectral function of the $\sigma$/$f_0(500)$ using the Sill parametrization according to \cite{GiacosaEtAl_2021}. Most importantly, with the chosen parametrization, the spectral function vanishes below the mass threshold given by twice the pion mass, thereby respecting kinematic constraints of the decay process -- a crucial improvement over the usual Breit-Wigner parametrization.}
    \label{fig:SillSpectralDensity}
\end{figure}

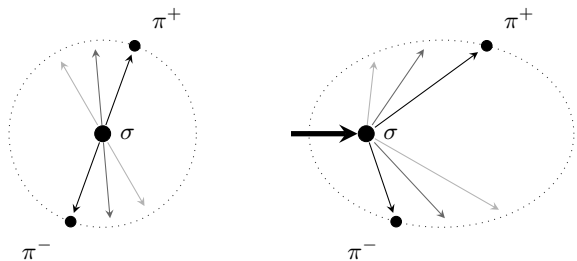
\begin{figure}
    \centering
    \begin{adjustbox}{valign=c}
        \begin{tikzpicture}
            \tikzmath{           
                \ma=2;
                \mb=0.7;
                \mc=0.7;
                \pa=0;
                \l={\ma^4+(\mb^2+\mc^2)^2+2*\ma^2*(\mb^2+\mc^2)};
                \p0={\pa*(\ma^2+\mb^2-\mc^2)/(2*\ma^2)};
                \atilde={sqrt(\ma^2+\pa^2)/(2*\ma^2)};
                \btilde={1/(2*\ma)};
                \a=sqrt(\l)*\atilde;
                \b=sqrt(\l)*\btilde;
                \myangle1 = 70;
                \myangle2 = 95;
                \myangle3 = 120;
            }     
            
            \node[label=right:$\sigma$] (sigma) at (0,0) {};
            \node[label=above right:$\pi^+$] (pi1) at ({\p0+\a*cos(\myangle1)},{\b*sin(\myangle1)}) {};
            \node (pi2) at ({\p0+\a*cos(\myangle2)},{\b*sin(\myangle2)}) {};
            \node (pi3) at ({\p0+\a*cos(\myangle3)},{\b*sin(\myangle3)}) {};
            \node[label=below left:$\pi^-$] (mpi1) at ({\p0+\a*cos(\myangle1+180)},{\b*sin(\myangle1+180)}) {};
            \node (mpi2) at ({\p0+\a*cos(\myangle2+180)},{\b*sin(\myangle2+180)}) {};
            \node (mpi3) at ({\p0+\a*cos(\myangle3+180)},{\b*sin(\myangle3+180)}) {};
            
            \draw[dotted,opacity=0.8] (\p0,0) ellipse ({\a} and {\b});
            
            \draw[fill] (sigma) circle (3pt);
            \draw[fill] (pi1) circle (2pt);  
            \draw[fill] (mpi1) circle (2pt);
            \draw[->,>=stealth] (sigma) -- (pi1);         
            \draw[->,>=stealth] (sigma) -- (mpi1);    
            
            \draw[opacity=0.6,->,>=stealth] (sigma) -- (pi2);    
            \draw[opacity=0.6,->,>=stealth] (sigma) -- (mpi2);    
            \draw[opacity=0.3,->,>=stealth] (sigma) -- (pi3);    
            \draw[opacity=0.3,->,>=stealth] (sigma) -- (mpi3);  
        \end{tikzpicture}
    \end{adjustbox}
    \hspace{1cm}
    \begin{adjustbox}{valign=c}
        \begin{tikzpicture}
            \tikzmath{           
                \ma=2;
                \mb=0.7;
                \mc=0.7;
                \pa=2;
                \l={\ma^4+(\mb^2+\mc^2)^2+2*\ma^2*(\mb^2+\mc^2)};
                \p0={\pa*(\ma^2+\mb^2-\mc^2)/(2*\ma^2)};
                \atilde={sqrt(\ma^2+\pa^2)/(2*\ma^2)};
                \btilde={1/(2*\ma)};
                \a=sqrt(\l)*\atilde;
                \b=sqrt(\l)*\btilde;
                \myangle1 = 70;
                \myangle2 = 95;
                \myangle3 = 120;
            }     
            
            \node[label=right:$\sigma$] (sigma) at (0,0) {};
            \node[label=above right:$\pi^+$] (pi1) at ({\p0+\a*cos(\myangle1)},{\b*sin(\myangle1)}) {};
            \node (pi2) at ({\p0+\a*cos(\myangle2)},{\b*sin(\myangle2)}) {};
            \node (pi3) at ({\p0+\a*cos(\myangle3)},{\b*sin(\myangle3)}) {};
            \node[label=below left:$\pi^-$] (mpi1) at ({\p0+\a*cos(\myangle1+180)},{\b*sin(\myangle1+180)}) {};
            \node (mpi2) at ({\p0+\a*cos(\myangle2+180)},{\b*sin(\myangle2+180)}) {};
            \node (mpi3) at ({\p0+\a*cos(\myangle3+180)},{\b*sin(\myangle3+180)}) {};
            
            \draw[dotted,opacity=0.8] (\p0,0) ellipse ({\a} and {\b});
            
            \draw[fill] (sigma) circle (3pt);
            \draw[fill] (pi1) circle (2pt);  
            \draw[fill] (mpi1) circle (2pt);
            \draw[->,>=stealth] (sigma) -- (pi1);         
            \draw[->,>=stealth] (sigma) -- (mpi1);    
            
            \draw[opacity=0.6,->,>=stealth] (sigma) -- (pi2);    
            \draw[opacity=0.6,->,>=stealth] (sigma) -- (mpi2);    
            \draw[opacity=0.3,->,>=stealth] (sigma) -- (pi3);    
            \draw[opacity=0.3,->,>=stealth] (sigma) -- (mpi3);  

            \node (pa0) at ({\pa/2},0) {};
            \draw[line width=2pt,<-,>=stealth] (sigma) -- ($(sigma)-(pa0)$);  
        \end{tikzpicture}
    \end{adjustbox}
    \caption{In a restframe of the heavier $\sigma$-resonance, we assume an isotropic distribution of the three-momenta of the decay products, with the allowed momenta being limited by three-momentum and energy conservation. In a frame where the resonance has non-vanishing three-momentum, the momentum distribution of the decay products is boosted accordingly. By this mechanism, the final momentum distribution of the decay products is a convolution of the momentum distribution of the primary resonance with the decay function \cite{SollfrankEtAl_1990, Mazeliauskas:2018irt}.}
    \label{fig:decaykinematics}
\end{figure}

The formalism presented above can, in principle, be used to study the particle production from all kinds of coherent sources. For a PRCC a subsequent decay of the unstable $\sigma/f_0(500)$ resonances into charged pions must be taken into account \citep{SollfrankEtAl_1990}. To that end, we replace $m^2$ with an integration parameter $\mu^2$ and weigh contributions by a spectral function $\rho(\mu^2)$. For the latter, we use the Sill parametrization \citep{GiacosaEtAl_2021},
\begin{equation}
    \rho(\mu^2) = - \text{Im}\left(\frac{1}{\mu^2-M^2 + i \Gamma \sqrt{\mu^2-4m_\pi^2} }\right) \theta(\mu^2-4m_\pi^2)\,,
\end{equation}
as depicted in Fig.~\ref{fig:SillSpectralDensity}, allowing a better treatment of the wide $\sigma$-resonance and the inclusion of threshhold effects compared to the usual Breit-Wigner parametrization, with $M$ and $\Gamma$ chosen such that this expression has a pole given by the parameters $\sqrt{s_\sigma} \simeq M_\sigma - i \Gamma_\sigma/2 = (449 - i 275)\; \mathrm{MeV}$ \cite{PELAEZ20161} in accordance with the recommended PDG values \cite{Navasothers_2024}. The decays are isotropic in the rest frame of the $\sigma/f_0(500)$ resonance. The momentum distribution of the resulting pions follows as a convolution of the momentum distribution of the initial resonances with an isotropic decay function, see illustration in Fig.~\ref{fig:decaykinematics}.

%Here we need as an input the $\sigma/f_0(500)$ spectral function, which we take from \citep{GiacosaEtAl_2021}. \textcolor{blue}{put values of mass and gamma, do we even need fig 2? it does not illustrate the new physics, just a boost in the rest frame; do we need more on the decay integral?}

The concrete dynamics leading to the PRCC and its relation to the dynamics of the QGP are so far not known from first principles, resulting in the functional form of the condensate field on the freeze-out hypersurface being unknown. In \cite{fsu_mods_00027527} we explored different parameterizations for $\sigma(t, \mathbf{x}) - f_\pi$ and observed a very limited influence on the final spectra, given relatively smooth parameterizations. Therefore, we now will assume for reasons of simplicity that the condensate field has a constant value $\Delta\sigma_{\text{coherent}}=35\;\mathrm{MeV}$ on the freeze-out hypersurface (defined by $T\approx 125 \; \mathrm{MeV}$).
% The resulting pion spectra are shown in Fig.~\ref{fig:ResultsALICE_PbPb2-76TeV_0-5} for collisions conducted at the LHC and at RHIC\todo{Qualitative agreement, quantitative discrepancies $\to$ fix order of this and the next paragraph, discuss Lu et al. curves, add temperature of hypersurface}.

%The results shown in \cref{fig:ResultsALICE_PbPb2-76TeV_0-5} correspond to a condensate amplitude of $\approx\SI{70}{\MeV}$ which is a plausible order of magnitude given the \gls{vev} of the \gls{cc}. There is an active debate on the nature of the $\sigma$-meson and its properties as one of the lightest resonances of \gls{qcd} [\citealp[Section~64]{Navasothers_2024}; \citealp{Pelaez_2016}], especially of the spectral distribution. A full discussion goes beyond the scope of the present analysis. We assume its properties to be dominated by those of the lightest available resonance in the Particle Data Book.
% \begin{widetext}
%%%\FloatBarrier
\begin{figure*}
    \centering
    \includegraphics[width=0.32\textwidth]{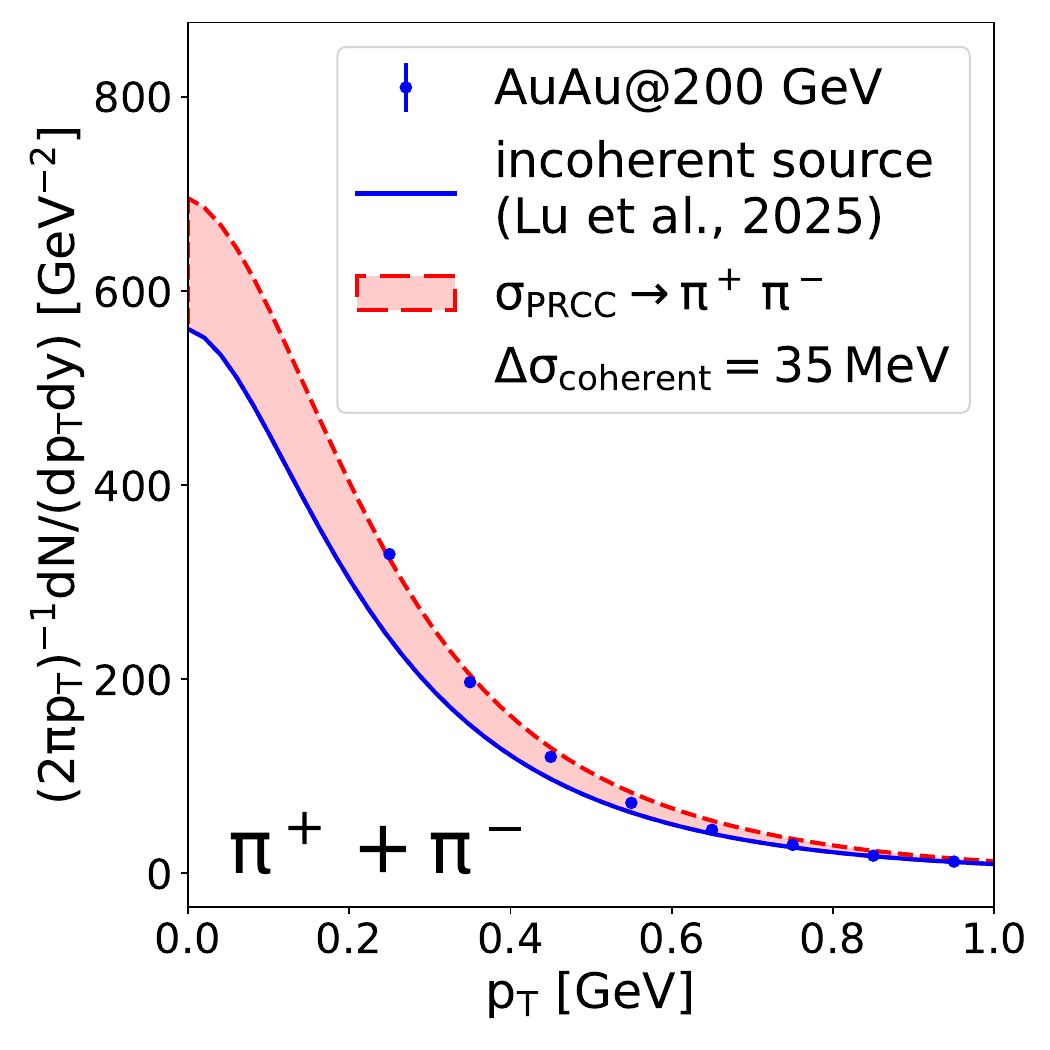}
    \includegraphics[width=0.32\textwidth]{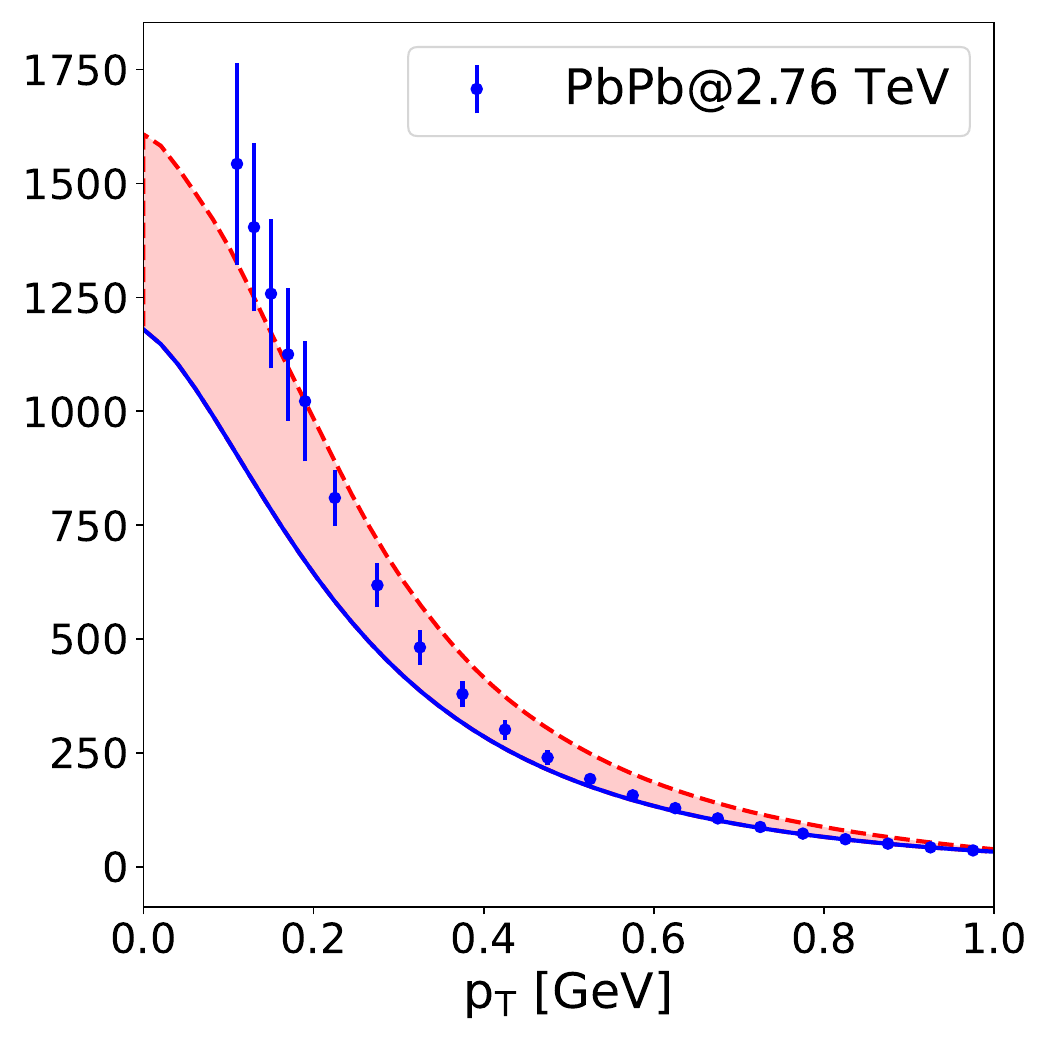}
    \includegraphics[width=0.32\textwidth]{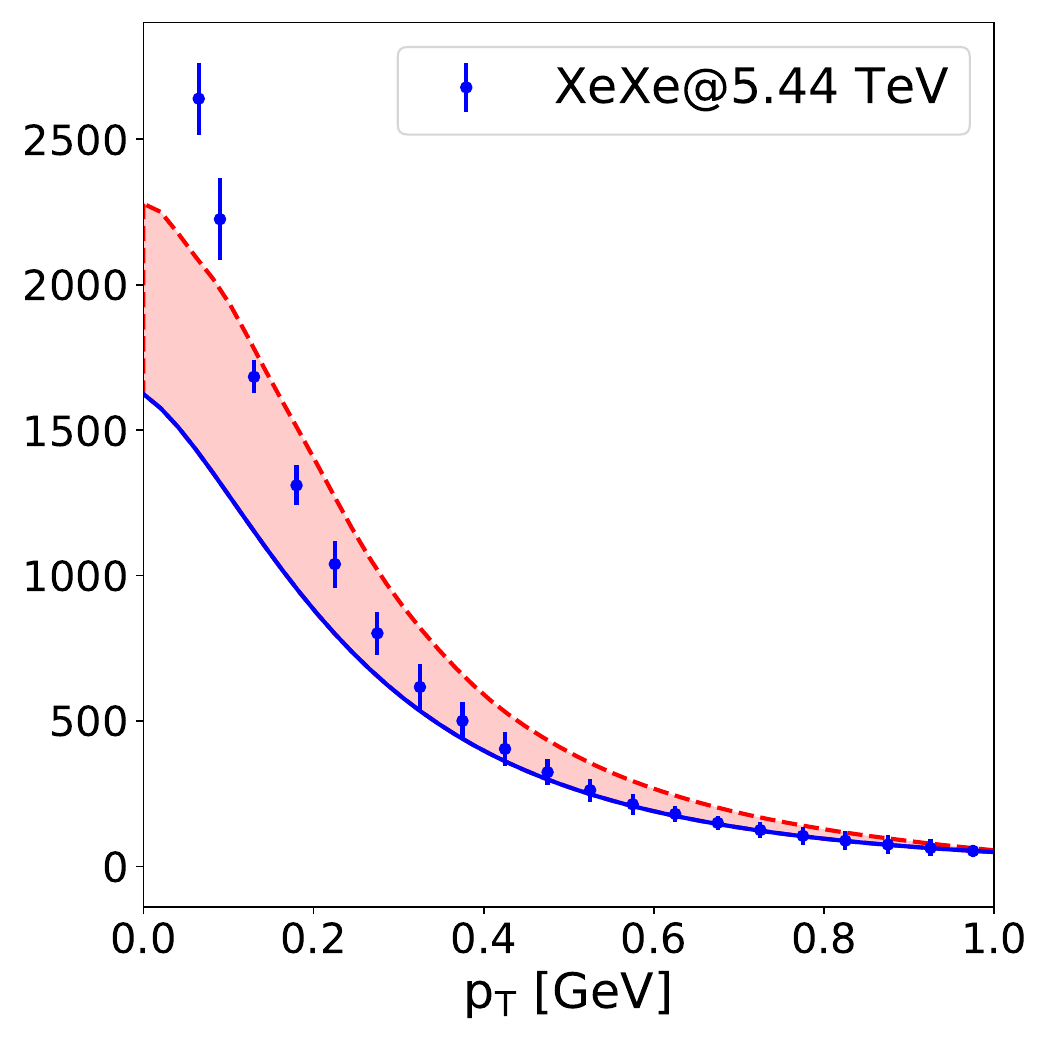}
    % \caption{Left: Sill parametrization \citep{GiacosaEtAl_2021} of the spectral function of the $f_0(500)$ with mass pole parameters $\sqrt{s_{\text{p}}}=M_{\text{p}}-\imagu\Gamma_{\text{p}}/2=400-\imagu\,200~\si{\MeV}$. Right: $\pi^\pm$-spectrum from fluid model Fluid$u$M \citep{FloerchingerEtAl_2019} plus decay of $\sigma$-condensate compared to ALICE data \citep{ALICECollaborationEtAl_2013}.}
    \caption{Comparison of pion spectra from experimental data \cite{PHENIX:2003iij,ALICE:2013mez,ALICE:2021lsv} with the fit result from \cite{LuEtAl_2025} and the contribution from the partially restored chiral condensate for different collision systems. We find a similar magnitude of pion enhancement at low momenta as found in the experimental results for all systems.}
    \label{fig:ResultsALICE_PbPb2-76TeV_0-5}
\end{figure*}
%%%\FloatBarrier
% \end{widetext}

%discussion of results, how many more pions do we get?
%AuAu: 217+70 (Fluidum+PRCC) --> 32.2\% 24.4\% (PRCC/Fluidum, PRCC/Total)
%PbPb: 523+224 (Fluidum+PRCC) --> 42.8\% 30\%
%XeXe: 733+338 (Fluidum+PRCC) --> 46.1\% 31.6\%

The resulting pion distributions for collisions conducted at the LHC and at RHIC, after performing the resonance decays of the $\sigma-$mesons, are depicted in Fig.~\ref{fig:ResultsALICE_PbPb2-76TeV_0-5} together with the experimental data \cite{PHENIX:2003iij,ALICE:2013mez,ALICE:2021lsv} and the pion distribution from incoherent sources \cite{LuEtAl_2025}. We find that our model shows a qualitative agreement with the experimentally found pion excess without requiring any fine-tuning of the parameters. In total, we observe the contribution of pions from the coherent source to be between $24\%$ to $32\%$ of the total pion multiplicity, depending on the collision system. While already yielding the right magnitude of pion excess, some discrepancies remain in the shapes of the spectra compared to the experiments; especially for the systems with larger collision energies, the experimentally observed enhancement peak is slightly narrower than our model prediction. %\sout{While already having the right magnitude, the discrepancies in the shapes of the spectra to the experimental results suggest}
This suggests that a simultaneous fit of the fluid-dynamic and PRCC model is necessary to find the global optimum in parameter space of the combined theory, given the data, also providing a more rigorous estimate of the PRCC amplitude in different collision systems. Such a simultaneous fit should also allow for a better description of the transition region in Fig.\ \ref{fig:ResultsALICE_PbPb2-76TeV_0-5}, in which the contribution of the condensate becomes less important. Additional uncertainties arise from the parameterization of the pole of the $\sigma/f_0(500)$ and the parameter values used therein.
%It is conceivable to obtain constraints for these parameters (mass and width of the $\sigma$-resonance) through calibration on experimental data, opening a novel way to study this particle.

\paragraph*{Conclusion and outlook}

In this work, we have demonstrated that a partially restored chiral condensate on the freeze-out surface has the potential to explain the abundance of pions produced over the expectation from thermal sources, measured across different colliders, collision systems, and energies. The partially restored condensate is hereby a result of the chiral phase transition, displacing the condensate field from its vacuum expectation value in the QGP fireball. We used this displacement to calculate the coherent spectrum of the $\sigma$-resonance on the freeze-out hypersurface, and in turn its contribution to the pion spectrum through resonance decays. We find that this contribution can explain the low-momentum pion excess qualitatively without fine-tuning of the parameters across different collision energies.

The inclusion of the PRCC in future global fits of experimental data will allow a more stringent test of our model and the determination of its parameters. In turn, these insights can lead to a deeper understanding of chiral symmetry, its restoration, and related dynamics. Through comparing data across different collision energies, a more precise determination of the pion excess and the condensate dynamics across different temperatures and baryon chemical potentials is possible. The recently conducted collisions of oxygen and neon at the LHC, together with the collisions of larger systems such as xenon and lead, also allow for a more thorough examination of the spatial extent and dynamics of the condensate field. 

The coupling of a dynamical condensate with the fluid model would allow for additional insights into the mechanism of chiral symmetry restoration, and a more concrete theoretical description, in future work.

The proposed model further has implications on momentum correlations in the final state, since an enhanced number of low momentum $\sigma$-particles in the intermediate state leads to large correlations between antipodally directed pions. Interestingly, in the classical source picture, the $\sigma$-particles emerging from the coherent state appear completely uncorrelated in Hanbury Brown-Twiss observables \cite{GyulassyEtAl_1979}. Further investigation is needed to estimate how the coherence assumption of the $\sigma$/$f_0(500)$ field affects final state momentum correlations of pions.

\paragraph*{Acknowledgments}
The authors would like to thank Francesco Giacosa for useful discussions. AK is supported by the U.S. Department of Energy, Office of Science, Office of Nuclear Physics, grant No. DE-FG02-05ER41367.

\bibliography{apssamp}% Produces the bibliography via BibTeX.

@article{Amelino-CameliaEtAl_1997,
  title = {Pion {{Production}} from {{Baked-Alaska Disoriented Chiral Condensate}}},
  author = {{Amelino-Camelia}, G. and Bjorken, J. D. and Larsson, S. E.},
  year = {1997},
  month = dec,
  journal = {Physical Review D},
  volume = {56},
  number = {11},
  eprint = {hep-ph/9706530},
  pages = {6942--6956},
  issn = {0556-2821, 1089-4918},
  doi = {10.1103/PhysRevD.56.6942},
  urldate = {2024-04-22},
  archiveprefix = {arXiv}
}

@article{AndersonEtAl_1995,
  title = {Observation of {{Bose-Einstein Condensation}} in a {{Dilute Atomic Vapor}}},
  author = {Anderson, M. H. and Ensher, J. R. and Matthews, M. R. and Wieman, C. E. and Cornell, E. A.},
  year = {1995},
  month = jul,
  journal = {Science},
  volume = {269},
  number = {5221},
  pages = {198--201},
  publisher = {American Association for the Advancement of Science},
  doi = {10.1126/science.269.5221.198},
  urldate = {2025-06-11}
}

@article{Anselm_1989,
  title = {Classical States of the Chiral Field and Nuclear Collisions at Very High Energy},
  author = {Anselm, A. A.},
  year = {1989},
  month = jan,
  journal = {Physics Letters B},
  volume = {217},
  number = {1},
  pages = {169--172},
  issn = {0370-2693},
  doi = {10.1016/0370-2693(89)91537-2},
  urldate = {2025-03-24}
}

@article{AokiEtAl_2006,
  title = {The Order of the Quantum Chromodynamics Transition Predicted by the Standard Model of Particle Physics},
  author = {Aoki, Y. and Endr{\H o}di, G. and Fodor, Z. and Katz, S. D. and Szab{\'o}, K. K.},
  year = {2006},
  month = oct,
  journal = {Nature},
  volume = {443},
  number = {7112},
  pages = {675--678},
  publisher = {Nature Publishing Group},
  issn = {1476-4687},
  doi = {10.1038/nature05120},
  urldate = {2025-06-12},
  copyright = {2006 Springer Nature Limited},
  langid = {english}
}

@article{AokiEtAl_2006a,
  title = {The {{QCD}} Transition Temperature: {{Results}} with Physical Masses in the Continuum Limit},
  shorttitle = {The {{QCD}} Transition Temperature},
  author = {Aoki, Y. and Fodor, Z. and Katz, S. D. and Szab{\'o}, K. K.},
  year = {2006},
  month = nov,
  journal = {Physics Letters B},
  volume = {643},
  number = {1},
  pages = {46--54},
  issn = {0370-2693},
  doi = {10.1016/j.physletb.2006.10.021},
  urldate = {2025-06-12}
}

@article{AokiEtAl_2009,
  title = {The {{QCD}} Transition Temperature: Results with Physical Masses in the Continuum Limit {{II}}.},
  shorttitle = {The {{QCD}} Transition Temperature},
  author = {Aoki, Y. and Borsanyi, Szabolcs and Durr, Stephan and Fodor, Zoltan and Katz, Sandor D. and Krieg, Stefan and Szabo, Kalman K.},
  year = {2009},
  journal = {JHEP},
  volume = {06},
  pages = {088},
  doi = {10.1088/1126-6708/2009/06/088}
}

@article{Arsene_2005,
  title = {Quark {{Gluon Plasma}} an {{Color Glass Condensate}} at {{RHIC}}? {{The}} Perspective from the {{BRAHMS}} Experiment},
  shorttitle = {Quark {{Gluon Plasma}} an {{Color Glass Condensate}} at {{RHIC}}?},
  author = {Arsene, I.},
  year = {2005},
  month = aug,
  journal = {Nuclear Physics A},
  volume = {757},
  number = {1-2},
  eprint = {nucl-ex/0410020},
  pages = {1--27},
  issn = {03759474},
  doi = {10.1016/j.nuclphysa.2005.02.130},
  urldate = {2025-03-17},
  archiveprefix = {arXiv}
}

@article{BackPHOBOSCollaboration_2005,
  title = {The {{PHOBOS Perspective}} on {{Discoveries}} at {{RHIC}}},
  author = {Back, B. B. and {PHOBOS Collaboration}},
  year = {2005},
  month = aug,
  journal = {Nuclear Physics A},
  volume = {757},
  number = {1-2},
  eprint = {nucl-ex/0410022},
  pages = {28--101},
  issn = {03759474},
  doi = {10.1016/j.nuclphysa.2005.03.084},
  urldate = {2025-03-17},
  archiveprefix = {arXiv}
}

@article{BegunFlorkowski_2015,
  title = {Bose-{{Einstein}} Condensation of Pions in Heavy-Ion Collisions at Energies Available at the {{CERN Large Hadron Collider}}},
  author = {Begun, Viktor and Florkowski, Wojciech},
  year = {2015},
  month = may,
  journal = {Physical Review C},
  volume = {91},
  number = {5},
  pages = {054909},
  publisher = {American Physical Society},
  doi = {10.1103/PhysRevC.91.054909},
  urldate = {2024-04-17}
}

@misc{Bjorken_1997,
  title = {Disoriented {{Chiral Condensate}}: {{Theory}} and {{Phenomenology}}},
  shorttitle = {Disoriented {{Chiral Condensate}}},
  author = {Bjorken, J. D.},
  year = {1997},
  month = dec,
  number = {arXiv:hep-ph/9712434},
  eprint = {hep-ph/9712434},
  publisher = {arXiv},
  doi = {10.48550/arXiv.hep-ph/9712434},
  urldate = {2024-04-17},
  archiveprefix = {arXiv}
}

@article{BradleyEtAl_1995,
  title = {Evidence of {{Bose-Einstein Condensation}} in an {{Atomic Gas}} with {{Attractive Interactions}}},
  author = {Bradley, C. C. and Sackett, C. A. and Tollett, J. J. and Hulet, R. G.},
  year = {1995},
  month = aug,
  journal = {Physical Review Letters},
  volume = {75},
  number = {9},
  pages = {1687--1690},
  publisher = {American Physical Society},
  doi = {10.1103/PhysRevLett.75.1687},
  urldate = {2025-06-11}
}

@article{CassingEtAl_2016,
  title = {Chiral Symmetry Restoration versus Deconfinement in Heavy-Ion Collisions at High Baryon Density},
  author = {Cassing, W. and Palmese, A. and Moreau, P. and Bratkovskaya, E. L.},
  year = {2016},
  month = jan,
  journal = {Physical Review C},
  volume = {93},
  number = {1},
  pages = {014902},
  publisher = {American Physical Society},
  doi = {10.1103/PhysRevC.93.014902},
  urldate = {2025-06-05}
}

@article{CooperFrye_1974,
  title = {Single-Particle Distribution in the Hydrodynamic and Statistical Thermodynamic Models of Multiparticle Production},
  author = {Cooper, Fred and Frye, Graham},
  year = {1974},
  month = jul,
  journal = {Physical Review D},
  volume = {10},
  number = {1},
  pages = {186--189},
  publisher = {American Physical Society},
  doi = {10.1103/PhysRevD.10.186},
  urldate = {2024-03-28}
}

@article{CornellWieman_2002,
  title = {Nobel {{Lecture}}: {{Bose-Einstein}} Condensation in a Dilute Gas, the First 70 Years and Some Recent Experiments},
  shorttitle = {Nobel {{Lecture}}},
  author = {Cornell, E. A. and Wieman, C. E.},
  year = {2002},
  month = aug,
  journal = {Reviews of Modern Physics},
  volume = {74},
  number = {3},
  pages = {875--893},
  publisher = {American Physical Society},
  doi = {10.1103/RevModPhys.74.875},
  urldate = {2025-06-11}
}

@article{DavisEtAl_1995,
  title = {Bose-{{Einstein Condensation}} in a {{Gas}} of {{Sodium Atoms}}},
  author = {Davis, K. B. and Mewes, M. -O. and Andrews, M. R. and {van Druten}, N. J. and Durfee, D. S. and Kurn, D. M. and Ketterle, W.},
  year = {1995},
  month = nov,
  journal = {Physical Review Letters},
  volume = {75},
  number = {22},
  pages = {3969--3973},
  publisher = {American Physical Society},
  doi = {10.1103/PhysRevLett.75.3969},
  urldate = {2025-06-11}
}

@article{FloerchingerEtAl_2019,
  title = {Fluid Dynamics of Heavy Ion Collisions with {{Mode}} Expansion ({{FluiduM}})},
  author = {Floerchinger, Stefan and Grossi, Eduardo and Lion, Jorrit},
  year = {2019},
  month = jul,
  journal = {Physical Review C},
  volume = {100},
  number = {1},
  eprint = {1811.01870},
  primaryclass = {hep-ph, physics:nucl-th},
  pages = {014905},
  issn = {2469-9985, 2469-9993},
  doi = {10.1103/PhysRevC.100.014905},
  urldate = {2024-08-29},
  archiveprefix = {arXiv}
}

@misc{FlorioEtAl_2025,
  title = {Supercooled {{Goldstones}} at the {{QCD}} Chiral Phase Transition},
  author = {Florio, Adrien and Grossi, Eduardo and Mazeliauskas, Aleksas and Soloviev, Alexander and Teaney, Derek},
  year = {2025},
  month = apr,
  number = {arXiv:2504.03516},
  eprint = {2504.03516},
  primaryclass = {hep-ph},
  publisher = {arXiv},
  doi = {10.48550/arXiv.2504.03516},
  urldate = {2025-04-14},
  archiveprefix = {arXiv}
}

@article{Gelis_2013,
  title = {Color {{Glass Condensate}} and {{Glasma}}},
  author = {Gelis, F.},
  year = {2013},
  month = jan,
  journal = {International Journal of Modern Physics A},
  volume = {28},
  number = {01},
  eprint = {1211.3327},
  primaryclass = {hep-ph},
  pages = {1330001},
  issn = {0217-751X, 1793-656X},
  doi = {10.1142/S0217751X13300019},
  urldate = {2025-04-07},
  archiveprefix = {arXiv}
}

@book{Gelis_2019,
  title = {Quantum {{Field Theory}}: {{From Basics}} to {{Modern Topics}}},
  shorttitle = {Quantum {{Field Theory}}},
  author = {Gelis, Fran{\c c}ois},
  year = {2019},
  month = jul,
  edition = {1},
  publisher = {Cambridge University Press},
  doi = {10.1017/9781108691550},
  urldate = {2025-03-25},
  copyright = {https://www.cambridge.org/core/terms},
  isbn = {978-1-108-69155-0 978-1-108-48090-1}
}

@article{GelisVenugopalan_2006,
  title = {Particle Production in Field Theories Coupled to Strong External Sources {{I}}. {{Formalism}} and Main Results},
  author = {Gelis, F. and Venugopalan, R.},
  year = {2006},
  month = oct,
  journal = {Nuclear Physics A},
  volume = {776},
  number = {3-4},
  eprint = {hep-ph/0601209},
  pages = {135--171},
  issn = {03759474},
  doi = {10.1016/j.nuclphysa.2006.07.020},
  urldate = {2025-02-20},
  archiveprefix = {arXiv}
}

@misc{GiacosaEtAl_2021,
  title = {A Simple Alternative to the {{Breit-Wigner}} Distribution},
  author = {Giacosa, Francesco and Okopi{\'n}ska, Anna and Shastry, Vanamali},
  year = {2021},
  month = nov,
  number = {arXiv:2106.03749},
  eprint = {2106.03749},
  publisher = {arXiv},
  doi = {10.48550/arXiv.2106.03749},
  urldate = {2024-11-27},
  archiveprefix = {arXiv}
}

@article{Glauber_1963,
  title = {The {{Quantum Theory}} of {{Optical Coherence}}},
  author = {Glauber, Roy J.},
  year = {1963},
  month = jun,
  journal = {Physical Review},
  volume = {130},
  number = {6},
  pages = {2529--2539},
  publisher = {American Physical Society},
  doi = {10.1103/PhysRev.130.2529},
  urldate = {2025-07-19}
}

@article{Goldstone_1961,
  title = {Field Theories with << {{Superconductor}} >> Solutions},
  author = {Goldstone, J.},
  year = {1961},
  month = jan,
  journal = {Il Nuovo Cimento (1955-1965)},
  volume = {19},
  number = {1},
  pages = {154--164},
  issn = {1827-6121},
  doi = {10.1007/BF02812722},
  urldate = {2025-03-24},
  langid = {english}
}

@article{GoldstoneEtAl_1962,
  title = {Broken {{Symmetries}}},
  author = {Goldstone, Jeffrey and Salam, Abdus and Weinberg, Steven},
  year = {1962},
  month = aug,
  journal = {Physical Review},
  volume = {127},
  number = {3},
  pages = {965--970},
  publisher = {American Physical Society},
  doi = {10.1103/PhysRev.127.965},
  urldate = {2025-03-24}
}

@article{GyulassyEtAl_1979,
  title = {Pion Interferometry of Nuclear Collisions. {{I}}. {{Theory}}},
  author = {Gyulassy, M. and Kauffmann, S. K. and Wilson, Lance W.},
  year = {1979},
  month = dec,
  journal = {Physical Review C},
  volume = {20},
  number = {6},
  pages = {2267--2292},
  publisher = {American Physical Society},
  doi = {10.1103/PhysRevC.20.2267},
  urldate = {2025-03-13}
}

@article{HotQCDCollaborationEtAl_2014,
  title = {{{QCD Phase Transition}} with {{Chiral Quarks}} and {{Physical Quark Masses}}},
  author = {{HotQCD Collaboration} and Bhattacharya, Tanmoy and Buchoff, Michael I. and Christ, Norman H. and Ding, H.-T. and Gupta, Rajan and Jung, Chulwoo and Karsch, F. and Lin, Zhongjie and Mawhinney, R. D. and McGlynn, Greg and Mukherjee, Swagato and Murphy, David and Petreczky, P. and Renfrew, Dwight and Schroeder, Chris and Soltz, R. A. and Vranas, P. M. and Yin, Hantao},
  year = {2014},
  month = aug,
  journal = {Physical Review Letters},
  volume = {113},
  number = {8},
  pages = {082001},
  publisher = {American Physical Society},
  doi = {10.1103/PhysRevLett.113.082001},
  urldate = {2025-06-13}
}

@article{Ketterle_2002,
  title = {Nobel Lecture: {{When}} Atoms Behave as Waves: {{Bose-Einstein}} Condensation and the Atom Laser},
  shorttitle = {Nobel Lecture},
  author = {Ketterle, Wolfgang},
  year = {2002},
  month = nov,
  journal = {Reviews of Modern Physics},
  volume = {74},
  number = {4},
  pages = {1131--1151},
  publisher = {American Physical Society},
  doi = {10.1103/RevModPhys.74.1131},
  urldate = {2025-06-11}
}

@article{LuEtAl_2025,
  title = {Quantification of the Low-{{pT}} Pion Excess in Heavy-Ion Collisions at the {{LHC}} and Top {{RHIC}} Energy},
  author = {Lu, P. and Kavak, R. and Dubla, A. and Masciocchi, S. and Selyuzhenkov, I.},
  year = 2025,
  month = jun,
  journal = {Nuclear Science and Techniques},
  volume = {36},
  number = {8},
  pages = {142},
  issn = {2210-3147},
  doi = {10.1007/s41365-025-01718-z}
}

@article{MohantySerreau_2005,
  title = {Disoriented {{Chiral Condensate}}: {{Theory}} and {{Experiment}}},
  shorttitle = {Disoriented {{Chiral Condensate}}},
  author = {Mohanty, Bedanga and Serreau, Julien},
  year = {2005},
  month = aug,
  journal = {Physics Reports},
  volume = {414},
  number = {6},
  eprint = {hep-ph/0504154},
  pages = {263--358},
  issn = {03701573},
  doi = {10.1016/j.physrep.2005.04.004},
  urldate = {2024-07-19},
  archiveprefix = {arXiv}
}

@article{MoreauEtAl_2017,
  title = {Evidence for Chiral Symmetry Restoration in Heavy-Ion Collisions},
  author = {Moreau, P. and Palmese, A. and Cassing, W. and Seifert, E. and Steinert, T. and Bratkovskaya, E. L.},
  year = {2017},
  month = nov,
  journal = {Nuclear Physics A},
  series = {The 26th {{International Conference}} on {{Ultra-relativistic Nucleus-Nucleus Collisions}}: {{Quark Matter}} 2017},
  volume = {967},
  pages = {836--839},
  issn = {0375-9474},
  doi = {10.1016/j.nuclphysa.2017.05.030},
  urldate = {2025-06-11}
}

@article{NambuJona-Lasinio_1961,
  title = {Dynamical {{Model}} of {{Elementary Particles Based}} on an {{Analogy}} with {{Superconductivity}}. {{I}}},
  author = {Nambu, Y. and {Jona-Lasinio}, G.},
  year = {1961},
  month = apr,
  journal = {Physical Review},
  volume = {122},
  number = {1},
  pages = {345--358},
  publisher = {American Physical Society},
  doi = {10.1103/PhysRev.122.345},
  urldate = {2025-03-20}
}

@article{Navasothers_2024,
  title = {Review of Particle Physics},
  author = {Navas, S. and others},
  year = {2024},
  journal = {Phys. Rev. D},
  volume = {110},
  number = {3},
  pages = {030001},
  doi = {10.1103/PhysRevD.110.030001}
}

@article{PalmeseEtAl_2016,
  title = {Chiral Symmetry Restoration in Heavy-Ion Collisions at Intermediate Energies},
  author = {Palmese, A. and Cassing, W. and Seifert, E. and Steinert, T. and Moreau, P. and Bratkovskaya, E. L.},
  year = {2016},
  month = oct,
  journal = {Physical Review C},
  volume = {94},
  number = {4},
  pages = {044912},
  publisher = {American Physical Society},
  doi = {10.1103/PhysRevC.94.044912},
  urldate = {2025-06-05}
}

@article{Pelaez_2016,
  shorttitle = {From Controversy to Precision on the Sigma Meson},
  author = {Pelaez, J. R.},
  year = {2016},
  month = nov,
  journal = {Physics Reports},
  volume = {658},
  eprint = {1510.00653},
  primaryclass = {hep-ph, physics:nucl-th},
  pages = {1--111},
  issn = {03701573},
  doi = {10.1016/j.physrep.2016.09.001},
  urldate = {2024-07-19},
  archiveprefix = {arXiv},
  title = {From Controversy to Precision on the Sigma Meson: A Review on the Status of the Non-Ordinary {$f_0(500)$} Resonance}
}

@article{PHENIXCollaborationAdcox_2005,
  title = {Formation of Dense Partonic Matter in Relativistic Nucleus-Nucleus Collisions at {{RHIC}}: {{Experimental}} Evaluation by the {{PHENIX}} Collaboration},
  shorttitle = {Formation of Dense Partonic Matter in Relativistic Nucleus-Nucleus Collisions at {{RHIC}}},
  author = {{PHENIX Collaboration} and Adcox, K.},
  year = {2005},
  month = aug,
  journal = {Nuclear Physics A},
  volume = {757},
  number = {1-2},
  eprint = {nucl-ex/0410003},
  pages = {184--283},
  issn = {03759474},
  doi = {10.1016/j.nuclphysa.2005.03.086},
  urldate = {2025-03-17},
  archiveprefix = {arXiv}
}

@article{PisarskiWilczek_1984,
  title = {Remarks on the Chiral Phase Transition in Chromodynamics},
  author = {Pisarski, Robert D. and Wilczek, Frank},
  year = {1984},
  month = jan,
  journal = {Physical Review D},
  volume = {29},
  number = {2},
  pages = {338--341},
  publisher = {American Physical Society},
  doi = {10.1103/PhysRevD.29.338},
  urldate = {2025-06-11}
}

@article{RajagopalWilczek_1993,
  title = {Emergence of Coherent Long Wavelength Oscillations after a Quench: Application to {{QCD}}},
  shorttitle = {Emergence of Coherent Long Wavelength Oscillations after a Quench},
  author = {Rajagopal, Krishna and Wilczek, Frank},
  year = {1993},
  month = sep,
  journal = {Nuclear Physics B},
  volume = {404},
  number = {3},
  pages = {577--589},
  issn = {0550-3213},
  doi = {10.1016/0550-3213(93)90591-C},
  urldate = {2025-01-27}
}

@incollection{RappEtAl_2010,
  title = {The {{Chiral Restoration Transition}} of {{QCD}} and {{Low Mass Dileptons}}},
  author = {Rapp, R. and Wambach, J. and van Hees, H.},
  year = {2010},
  volume = {23},
  eprint = {0901.3289},
  primaryclass = {hep-ph},
  pages = {134--175},
  doi = {10.1007/978-3-642-01539-7_6},
  urldate = {2025-06-12},
  archiveprefix = {arXiv}
}

@article{RappvanHees_2016,
  title = {Thermal Dileptons as Fireball Thermometer and Chronometer},
  author = {Rapp, Ralf and {van Hees}, Hendrik},
  year = {2016},
  month = feb,
  journal = {Physics Letters B},
  volume = {753},
  pages = {586--590},
  issn = {0370-2693},
  doi = {10.1016/j.physletb.2015.12.065},
  urldate = {2025-06-11}
}

@article{SollfrankEtAl_1990,
  title = {The Influence of Resonance Decays on the {\emph{p}}{{T}} Spectra from Heavy-Ion Collisions},
  author = {Sollfrank, Josef and Koch, Peter and Heinz, Ulrich},
  year = {1990},
  month = dec,
  journal = {Physics Letters B},
  volume = {252},
  number = {2},
  pages = {256--264},
  issn = {0370-2693},
  doi = {10.1016/0370-2693(90)90870-C},
  urldate = {2025-04-14}
}

@article{STARCollaborationAdams_2005,
  title = {Experimental and {{Theoretical Challenges}} in the {{Search}} for the {{Quark Gluon Plasma}}: {{The STAR Collaboration}}'s {{Critical Assessment}} of the {{Evidence}} from {{RHIC Collisions}}},
  shorttitle = {Experimental and {{Theoretical Challenges}} in the {{Search}} for the {{Quark Gluon Plasma}}},
  author = {{STAR Collaboration} and Adams, J.},
  year = {2005},
  month = aug,
  journal = {Nuclear Physics A},
  volume = {757},
  number = {1-2},
  eprint = {nucl-ex/0501009},
  pages = {102--183},
  issn = {03759474},
  doi = {10.1016/j.nuclphysa.2005.03.085},
  urldate = {2025-03-17},
  archiveprefix = {arXiv}
}

@article{SungEtAl_2021,
  title = {{\emph{K}}1/{{{\emph{K}}}}⁎ Enhancement as a Signature of Chiral Symmetry Restoration in Heavy Ion Collisions},
  author = {Sung, Hae-Som and Cho, Sungtae and Hong, Juhee and Lee, Su Houng and Lim, Sanghoon and Song, Taesoo},
  year = {2021},
  month = aug,
  journal = {Physics Letters B},
  volume = {819},
  pages = {136388},
  issn = {0370-2693},
  doi = {10.1016/j.physletb.2021.136388},
  urldate = {2025-06-05}
}

@misc{ZhouEtAl_2024,
  title = {Effects of Chiral Symmetry Restoration on Dilepton Production in Heavy Ion Collisions},
  author = {Zhou, Wen-Hao and Ko, Che Ming and Sun, Kai-Jia},
  year = {2024},
  month = dec,
  number = {arXiv:2412.18895},
  eprint = {2412.18895},
  primaryclass = {nucl-th},
  publisher = {arXiv},
  doi = {10.48550/arXiv.2412.18895},
  urldate = {2025-06-05},
  archiveprefix = {arXiv}
}

@article{ALICE:2022wpn,
    author = "Acharya, Shreyasi and others",
    collaboration = "ALICE",
    title = "{The ALICE experiment: a journey through QCD}",
    eprint = "2211.04384",
    archivePrefix = "arXiv",
    primaryClass = "nucl-ex",
    reportNumber = "CERN-EP-2022-227",
    doi = "10.1140/epjc/s10052-024-12935-y",
    journal = "Eur. Phys. J. C",
    volume = "84",
    number = "8",
    pages = "813",
    year = "2024"
}

@article{Foka:2016vta,
    author = "Foka, Panagiota and Janik, Ma{\l}gorzata Anna",
    title = "{An overview of experimental results from ultra-relativistic heavy-ion collisions at the CERN LHC: Bulk properties and dynamical evolution}",
    eprint = "1702.07233",
    archivePrefix = "arXiv",
    primaryClass = "hep-ex",
    doi = "10.1016/j.revip.2016.11.002",
    journal = "Rev. Phys.",
    volume = "1",
    pages = "154--171",
    year = "2016"
}

@article{Andronic:2014zha,
    author = "Andronic, Anton",
    title = "{An overview of the experimental study of quark-gluon matter in high-energy nucleus-nucleus collisions}",
    eprint = "1407.5003",
    archivePrefix = "arXiv",
    primaryClass = "nucl-ex",
    doi = "10.1142/S0217751X14300476",
    journal = "Int. J. Mod. Phys. A",
    volume = "29",
    pages = "1430047",
    year = "2014"
}

@article{Schukraft:2013wba,
    author = "Schukraft, J.",
    editor = {Ekel{\"o}f, Tord},
    title = "{Heavy ion physics at the Large Hadron Collider: what is new? What is next?}",
    eprint = "1311.1429",
    archivePrefix = "arXiv",
    primaryClass = "hep-ex",
    doi = "10.1088/0031-8949/2013/T158/014003",
    journal = "Phys. Scripta T",
    volume = "158",
    pages = "014003",
    year = "2013"
}

@article{Muller:2013dea,
    author = {M{\"u}ller, Berndt},
    editor = {Ekel{\"o}f, Tord},
    title = "{Investigation of Hot QCD Matter: Theoretical Aspects}",
    eprint = "1309.7616",
    archivePrefix = "arXiv",
    primaryClass = "nucl-th",
    doi = "10.1088/0031-8949/2013/T158/014004",
    journal = "Phys. Scripta T",
    volume = "158",
    pages = "014004",
    year = "2013"
}

@article{Busza:2018rrf,
    author = "Busza, Wit and Rajagopal, Krishna and van der Schee, Wilke",
    title = "{Heavy Ion Collisions: The Big Picture, and the Big Questions}",
    eprint = "1802.04801",
    archivePrefix = "arXiv",
    primaryClass = "hep-ph",
    reportNumber = "MIT-CTP-4892, MIT-CTP/4892",
    doi = "10.1146/annurev-nucl-101917-020852",
    journal = "Ann. Rev. Nucl. Part. Sci.",
    volume = "68",
    pages = "339--376",
    year = "2018"
}

@book{Romatschke:2017ejr,
    author = "Romatschke, Paul and Romatschke, Ulrike",
    title = "{Relativistic Fluid Dynamics In and Out of Equilibrium}",
    eprint = "1712.05815",
    archivePrefix = "arXiv",
    primaryClass = "nucl-th",
    doi = "10.1017/9781108651998",
    isbn = "978-1-108-48368-1, 978-1-108-75002-8",
    publisher = "Cambridge University Press",
    series = "Cambridge Monographs on Mathematical Physics",
    month = "5",
    year = "2019"
}

@article{Soloviev:2021lhs,
    author = "Soloviev, Alexander",
    title = "{Hydrodynamic attractors in heavy ion collisions: a review}",
    eprint = "2109.15081",
    archivePrefix = "arXiv",
    primaryClass = "hep-th",
    doi = "10.1140/epjc/s10052-022-10282-4",
    journal = "Eur. Phys. J. C",
    volume = "82",
    number = "4",
    pages = "319",
    year = "2022"
}

@article{Jaiswal:2016hex,
    author = "Jaiswal, Amaresh and Roy, Victor",
    title = "{Relativistic hydrodynamics in heavy-ion collisions: general aspects and recent developments}",
    eprint = "1605.08694",
    archivePrefix = "arXiv",
    primaryClass = "nucl-th",
    doi = "10.1155/2016/9623034",
    journal = "Adv. High Energy Phys.",
    volume = "2016",
    pages = "9623034",
    year = "2016"
}

@article{Shen:2020mgh,
    author = "Shen, Chun and Yan, Li",
    title = "{Recent development of hydrodynamic modeling in heavy-ion collisions}",
    eprint = "2010.12377",
    archivePrefix = "arXiv",
    primaryClass = "nucl-th",
    doi = "10.1007/s41365-020-00829-z",
    journal = "Nucl. Sci. Tech.",
    volume = "31",
    number = "12",
    pages = "122",
    month = "10",
    year = "2020"
}

@article{ISRAEL1979341,
title = {Transient relativistic thermodynamics and kinetic theory},
journal = {Annals of Physics},
volume = {118},
number = {2},
pages = {341-372},
year = {1979},
issn = {0003-4916},
doi = {https://doi.org/10.1016/0003-4916(79)90130-1},
url = {https://www.sciencedirect.com/science/article/pii/0003491679901301},
author = {W. Israel and J.M. Stewart}
}

@article{Moreland:2014oya,
    author = "Moreland, J. Scott and Bernhard, Jonah E. and Bass, Steffen A.",
    title = "{Alternative ansatz to wounded nucleon and binary collision scaling in high-energy nuclear collisions}",
    eprint = "1412.4708",
    archivePrefix = "arXiv",
    primaryClass = "nucl-th",
    doi = "10.1103/PhysRevC.92.011901",
    journal = "Phys. Rev. C",
    volume = "92",
    number = "1",
    pages = "011901",
    year = "2015"
}

@phdthesis{Moreland:2019szz,
    author = "Moreland, J. Scott",
    title = "{Initial conditions of bulk matter in ultrarelativistic nuclear collisions}",
    eprint = "1904.08290",
    archivePrefix = "arXiv",
    primaryClass = "nucl-th",
    school = "Duke U.",
    year = "2019"
}

@article{Floerchinger:2013hza,
    author = "Floerchinger, Stefan and Wiedemann, Urs Achim",
    title = "{Kinetic freeze-out, particle spectra and harmonic flow coefficients from mode-by-mode hydrodynamics}",
    eprint = "1311.7613",
    archivePrefix = "arXiv",
    primaryClass = "hep-ph",
    reportNumber = "CERN-PH-TH-2013-288",
    doi = "10.1103/PhysRevC.89.034914",
    journal = "Phys. Rev. C",
    volume = "89",
    number = "3",
    pages = "034914",
    year = "2014"
}

@article{Floerchinger:2014fta,
    author = "Floerchinger, Stefan and Wiedemann, Urs Achim",
    title = "{Statistics of initial density perturbations in heavy ion collisions and their fluid dynamic response}",
    eprint = "1405.4393",
    archivePrefix = "arXiv",
    primaryClass = "hep-ph",
    reportNumber = "CERN-PH-TH-2014-092",
    doi = "10.1007/JHEP08(2014)005",
    journal = "JHEP",
    volume = "08",
    pages = "005",
    year = "2014"
}

@article{Bebie:1991ij,
    author = "Bebie, H. and Gerber, P. and Goity, J. L. and Leutwyler, H.",
    title = "{The Role of the entropy in an expanding hadronic gas}",
    reportNumber = "BUTP-91-14-BERN, PSI-PR-91-09",
    doi = "10.1016/0550-3213(92)90005-V",
    journal = "Nucl. Phys. B",
    volume = "378",
    pages = "95--128",
    year = "1992"
}

@article{Kirchner:2023fsj,
    author = "Kirchner, Andreas and Grossi, Eduardo and Floerchinger, Stefan",
    title = "{Cooper-Frye spectra of hadrons with viscous corrections including feed down from resonance decays}",
    eprint = "2308.10616",
    archivePrefix = "arXiv",
    primaryClass = "hep-ph",
    journal = "",
    month = "8",
    year = "2023"
}

@article{Mazeliauskas:2018irt,
    author = "Mazeliauskas, Aleksas and Floerchinger, Stefan and Grossi, Eduardo and Teaney, Derek",
    title = "{Fast resonance decays in nuclear collisions}",
    eprint = "1809.11049",
    archivePrefix = "arXiv",
    primaryClass = "nucl-th",
    doi = "10.1140/epjc/s10052-019-6791-7",
    journal = "Eur. Phys. J. C",
    volume = "79",
    number = "3",
    pages = "284",
    year = "2019"
}

@article{PhysRevC.101.014910,
  title = {Temperature and fluid velocity on the freeze-out surface from $\ensuremath{\pi}$, $K$, and $p$ spectra in $pp$, $p$-Pb, and Pb-Pb collisions},
  author = {Mazeliauskas, Aleksas and Vislavicius, Vytautas},
  journal = {Phys. Rev. C},
  volume = {101},
  issue = {1},
  pages = {014910},
  numpages = {8},
  year = {2020},
  month = {Jan},
  publisher = {American Physical Society},
  doi = {10.1103/PhysRevC.101.014910},
  url = {https://link.aps.org/doi/10.1103/PhysRevC.101.014910}
}

@article{Vermunt:2023zsk,
    author = "Vermunt, L. and Seemann, Y. and Dubla, A. and Floerchinger, S. and Grossi, E. and Kirchner, A. and Masciocchi, S. and Selyuzhenkov, I.",
    title = "{Mapping properties of the quark gluon plasma in Pb-Pb and Xe-Xe collisions at energies available at the CERN Large Hadron Collider}",
    eprint = "2308.16722",
    archivePrefix = "arXiv",
    primaryClass = "hep-ph",
    doi = "10.1103/PhysRevC.108.064908",
    journal = "Phys. Rev. C",
    volume = "108",
    number = "6",
    pages = "064908",
    year = "2023"
}

@article{Nijs:2020roc,
    author = {Nijs, Govert and van der Schee, Wilke and G{\"u}rsoy, Umut and Snellings, Raimond},
    title = "{Bayesian analysis of heavy ion collisions with the heavy ion computational framework Trajectum}",
    eprint = "2010.15134",
    archivePrefix = "arXiv",
    primaryClass = "nucl-th",
    reportNumber = "CERN-TH-2020-175, MIT-CTP/5251",
    doi = "10.1103/PhysRevC.103.054909",
    journal = "Phys. Rev. C",
    volume = "103",
    number = "5",
    pages = "054909",
    year = "2021"
}

@article{JETSCAPE:2023nuf,
    author = "Mankolli, Andi and others",
    collaboration = "JETSCAPE",
    title = "{3D multi-system Bayesian calibration with energy conservation to study rapidity-dependent dynamics of nuclear collisions}",
    eprint = "2401.00402",
    archivePrefix = "arXiv",
    primaryClass = "nucl-th",
    doi = "10.1051/epjconf/202429605010",
    journal = "EPJ Web Conf.",
    volume = "296",
    pages = "05010",
    year = "2024"
}

@phdthesis{Bernhard:2018hnz,
    author = "Bernhard, Jonah E.",
    title = "{Bayesian parameter estimation for relativistic heavy-ion collisions}",
    eprint = "1804.06469",
    archivePrefix = "arXiv",
    primaryClass = "nucl-th",
    school = "Duke U.",
    month = "4",
    year = "2018"
}

@article{Paquet:2023rfd,
    author = "Paquet, Jean-Fran{\c{c}}ois",
    title = "{Applications of emulation and Bayesian methods in heavy-ion physics}",
    eprint = "2310.17618",
    archivePrefix = "arXiv",
    primaryClass = "nucl-th",
    doi = "10.1088/1361-6471/ad6a2b",
    journal = "J. Phys. G",
    volume = "51",
    number = "10",
    pages = "103001",
    year = "2024"
}

@article{Anselm:1991pi,
    author = "Anselm, A. A. and Ryskin, M. G.",
    title = "{Production of classical pion field in heavy ion high-energy collisions}",
    doi = "10.1016/0370-2693(91)91073-5",
    journal = "Phys. Lett. B",
    volume = "266",
    pages = "482--484",
    year = "1991"
}

@article{Blaizot:1992at,
    author = "Blaizot, Jean-Paul and Krzywicki, Andre",
    title = "{Soft pion emission in high-energy heavy ion collisions}",
    reportNumber = "LPTHE-ORSAY-92-11",
    doi = "10.1103/PhysRevD.46.246",
    journal = "Phys. Rev. D",
    volume = "46",
    pages = "246--251",
    year = "1992"
}

@article{Savchuk:2022aev,
    author = "Savchuk, Oleh and Motornenko, Anton and Steinheimer, Jan and Vovchenko, Volodymyr and Bleicher, Marcus and Gorenstein, Mark and Galatyuk, Tetyana",
    title = "{Enhanced dilepton emission from a phase transition in dense matter}",
    eprint = "2209.05267",
    archivePrefix = "arXiv",
    primaryClass = "nucl-th",
    doi = "10.1088/1361-6471/acfccf",
    journal = "J. Phys. G",
    volume = "50",
    number = "12",
    pages = "125104",
    year = "2023"
}

@article{Seck:2020qbx,
    author = "Seck, Florian and Galatyuk, Tetyana and Mukherjee, Ayon and Rapp, Ralf and Steinheimer, Jan and Stroth, Joachim and Wiest, Maximilian",
    title = "{Dilepton signature of a first-order phase transition}",
    eprint = "2010.04614",
    archivePrefix = "arXiv",
    primaryClass = "nucl-th",
    doi = "10.1103/PhysRevC.106.014904",
    journal = "Phys. Rev. C",
    volume = "106",
    number = "1",
    pages = "014904",
    year = "2022"
}

@book{Donoghue:1992dd,
    author = "Donoghue, J. F. and Golowich, E. and Holstein, Barry R.",
    title = "{Dynamics of the standard model}",
    doi = "10.1017/CBO9780511524370",
    publisher = "CUP",
    volume = "2",
    year = "2014"
}

@book{Schleich2001,
  title="Quantum Optics in Phase Space",
  author="Schleich, Wolfgang P.",
  publisher="Wiley-VCH",
  address="Berlin",
  year="2001"
}

@book{Altland_Simons_2010, 
place={Cambridge},
edition={2},
title={Condensed Matter Field Theory},
publisher={Cambridge University Press},
author={Altland, Alexander and Simons, Ben D.},
year={2010}}

@article{Florio:2025lvu,
    author = "Florio, Adrien and Grossi, Eduardo and Mazeliauskas, Aleksas and Soloviev, Alexander and Teaney, Derek",
    title = "{Quenching through the QCD chiral phase transition}",
    eprint = "2504.03514",
    archivePrefix = "arXiv",
    primaryClass = "hep-lat",
    journal = "",
    month = "4",
    year = "2025"
}

@article{Broniowski:2015oha,
    author = "Broniowski, Wojciech and Giacosa, Francesco and Begun, Viktor",
    title = "{Cancellation of the $\sigma$ meson in thermal models}",
    eprint = "1506.01260",
    archivePrefix = "arXiv",
    primaryClass = "nucl-th",
    doi = "10.1103/PhysRevC.92.034905",
    journal = "Phys. Rev. C",
    volume = "92",
    number = "3",
    pages = "034905",
    year = "2015"
}

@article{Andronic:2008gu,
    author = "Andronic, A. and Braun-Munzinger, P. and Stachel, J.",
    title = "{Thermal hadron production in relativistic nuclear collisions: The Hadron mass spectrum, the horn, and the QCD phase transition}",
    eprint = "0812.1186",
    archivePrefix = "arXiv",
    primaryClass = "nucl-th",
    doi = "10.1016/j.physletb.2009.06.021",
    journal = "Phys. Lett. B",
    volume = "673",
    pages = "142--145",
    year = "2009",
    note = "[Erratum: Phys.Lett.B 678, 516 (2009)]"
}

@article{PELAEZ20161,
title = {From controversy to precision on the sigma meson: A review on the status of the non-ordinary f0(500) resonance},
journal = {Physics Reports},
volume = {658},
pages = {1-111},
year = {2016},
note = {From controversy to precision on the sigma meson: A review on the status of the non-ordinary f0(500) resonance},
issn = {0370-1573},
doi = {https://doi.org/10.1016/j.physrep.2016.09.001},
url = {https://www.sciencedirect.com/science/article/pii/S0370157316302952},
author = {José R. Peláez}
}

@mastersthesis{fsu_mods_00027527,
  author = 	{Bruschke, Tobias},
  title = 	{Disoriented chiral condensate effects on soft pion spectra in heavy ion collisions},
  year = 	{2025},
  school = 	{Friedrich-Schiller-Universit{\"a}t Jena},
  doi = 	{10.22032/dbt.67475},
  url = 	{https://uri.gbv.de/document/opac-de-27:ppn:1936049716},
  url = 	{https://doi.org/10.22032/dbt.67475},
  language = 	{english}
}

@book{Rezzolla:2013dea,
    author = "Rezzolla, Luciano and Zanotti, Olindo",
    title = "{Relativistic Hydrodynamics}",
    doi = "10.1093/acprof:oso/9780198528906.001.0001",
    isbn = "978-0-19-174650-5, 978-0-19-852890-6",
    publisher = "Oxford University Press",
    month = "9",
    year = "2013"
}

@book{Landau1987Fluid,
  author = {Landau, L. D. and Lifshitz, E. M.},
  day = 15,
  edition = 2,
  howpublished = {Paperback},
  isbn = {0750627670},
  month = jan,
  publisher = {Butterworth-Heinemann},
  series = {Course of theoretical physics / by L. D. Landau and E. M. Lifshitz, Vol. 6},
  title = {Fluid Mechanics, Second Edition: Volume 6 (Course of Theoretical Physics)},
  url = {http://www.worldcat.org/isbn/0750627670},
  year = 1987
}

@article{ALICE:2013mez,
    author = "Abelev, Betty and others",
    collaboration = "ALICE",
    title = "{Centrality dependence of $\pi$, K, p production in Pb-Pb collisions at $\sqrt{s_{NN}}$ = 2.76 TeV}",
    eprint = "1303.0737",
    archivePrefix = "arXiv",
    primaryClass = "hep-ex",
    reportNumber = "CERN-PH-EP-2013-019",
    doi = "10.1103/PhysRevC.88.044910",
    journal = "Phys. Rev. C",
    volume = "88",
    pages = "044910",
    year = "2013"
}

\end{document}